# Substrate effects in high gain, low operating voltage SnSe$_2$ photoconductor

Krishna Murali, Sangeeth Kallatt, and Kausik Majumdar[*]

Department of Electrical Communication Engineering, Indian Institute of Science, Bangalore 560012, India

[*]*Corresponding author, email: kausikm@iisc.ac.in*

**ABSTRACT:** High gain photoconductive devices find wide spread applications in low intensity light detection. Ultra-thin layered materials have recently attracted a lot of attention from researchers in this regard. However, in general, a large operating voltage is required to obtain large responsivity in these devices. In addition, the characteristics are often confounded by substrate induced trap effects. Here we report multi-layer SnSe$_2$ based photoconductive devices using two different structures: (1) SiO$_2$ substrate supported interdigitated electrode (IDE), and (2) suspended channel. The IDE device exhibits a responsivity of $\approx 10^3$ A/W and $\approx 8.66 \times 10^4$ A/W at operating voltages of 1 mV and 100 mV, respectively – a superior low voltage performance over existing literature on planar 2D structures. However, the responsivity reduces by more than two orders of magnitude, while the transient response improves for the suspended device – providing insights into the critical role played by the channel-substrate interface in the gain mechanism. The results, on one hand, are promising for highly sensitive photoconductive applications consuming ultra-low power, and on the other hand, show a generic methodology that could be applied to other layered material based photoconductive devices as well for extracting the intrinsic behavior.

## 1. Introduction:

Ultra-thin layered materials are promising candidates for low cost, flexible electronic and opto-electronic applications[1]. While transition metal dichalcogenides (for example, sulfides and selenides of Mo and W) have attracted a lot of attention, other layered materials including group IIIA-VIA (sulfides and selenides of Ga and In) and group IVA-VIA (sulfides and selenides of Ge, Sn) compounds also exhibit excellent opto-electronic properties[2]. $SnSe_2$ is one such layered material, which has a very similar structure as that of transition metal dichalcogenides. In a monolayer, Sn atoms are sandwiched between the Se atoms, while multiple layers are weakly coupled through van der Waals interaction in the out of plane direction. $SnSe_2$ has been reported to be an air stable material, and can be deposited using MBE growth[3], spray pyrolysis[4] and mechanical exfoliation[5,6,7]. While several works report moderate carrier mobility[5,7] and weak gate tunability due to inherent large doping[7], large gate modulation of conductivity has also been reported[5].

$SnSe_2$ based photodetectors have been recently reported[8,7,9]. However, most of the photodetector devices reported on $SnSe_2$, and on layered materials in general, use large operating voltage to achieve better sensitivity – a major limitation for low power applications. Strong light absorption[10], coupled with excellent carrier transport properties of $SnSe_2$ provide an opportunity for building highly sensitive photodetection devices operating at small bias voltage – a regime which has not yet been explored. In addition, the photo response of layered material based devices is often dominated by carrier trapping at the channel-substrate interface. In this work, by using high quality contacts to multi-layer $SnSe_2$ in an inter-digitated electrode (IDE) structure, we achieve responsivity of $\approx 10^3$ A/W at 1 mV external bias, and the responsivity increases to $\approx 8.66 \times 10^4$ A/W at 100 mV. To investigate the origin of such high gain, we also compare the results with that of a suspended $SnSe_2$ device where the transient response is improved at the expense of suppressed gain – providing insights into the gain mechanism.

## 2. Experiment

### 2.1 Characterization of SnSe$_2$ flakes

Thickness of the exfoliated flakes was confirmed with Atomic Force Microscopy (AFM), as shown in Fig. 1a. The exfoliated crystals exhibit strong thickness dependence in Raman scattered signals[11,12], as illustrated in Fig. 1b. Two prominent peaks, namely E$_g$ and A$_{1g}$, are clearly observed. With an increase in thickness, the separation (Δ) between A$_{1g}$ and E$_g$ peaks reduces, particularly when the thickness is less than 8 nm (inset of Fig. 1b), and can be used to identify thickness of SnSe$_2$ flakes. On the other hand, with an increase in laser power, we observe that both the peaks exhibit strong shift towards lower wavenumber, as shown in Fig. 1c. Here zero corresponds to peak position at very small laser power. Such a shift can be attributed to the anharmonic effect[13] of SnSe$_2$ phonons resulting from an enhancement in local temperature of the sample due to increasing laser power[14].

### 2.2 Fabrication of Substrate supported and suspended structure

Thin layers of SnSe$_2$ are exfoliated on 285 nm thick SiO$_2$ covered Si substrate. The flakes are identified by the observation under optical microscopy, followed by Raman spectroscopy. SnSe$_2$ flakes having thickness of 10-15 nm are selected to fabricate photoconductor using inter-digitated electrode (IDE) structure, as schematically shown in Fig. 2a. The electrodes are defined using electron beam lithography followed by the deposition of Ni (10 nm)/Au (50 nm) using electron beam evaporation technique, and subsequent lift off. The top inset of Fig. 2b shows the optical image of the fabricated device.

For the suspended structure, first closely spaced metal pads are patterned by optical lithography, followed by deposition of Ni (10 nm)/Au (50 nm) using electron beam evaporation technique, and subsequently lift-off. The SnSe$_2$ flakes are then exfoliated on top of these metal pads carefully such that the flake is completely suspended without touching the SiO$_2$ substrate. The schematic of the suspended device is depicted in Fig. 2c. The cross section scanning electron micrograph (SEM) of the suspended device is shown in the inset of Fig. 2d and clearly indicates that the SnSe$_2$ film is completely isolated from the substrate.

## 3. Results and Discussions:

### 3.1 Electrical transport properties of SnSe₂ under dark condition

The black triangles in Fig. 2b indicate the dark current ($I_{dark}$) of a representative substrate supported IDE device. The strong linearity suggests good quality ohmic contacts with a dark current of 0.56 µA/µm at $V_{ds} = 1$ mV. On the other hand, the SnSe₂ channel being completely isolated from the substrate in the suspended structure, is expected to exhibit intrinsic behaviour of SnSe₂ as a photoconductor. The dark current of the suspended device, as shown by the black triangles in Fig. 2d, exhibits excellent linear characteristics as well. This indicated good contact quality for both the structures.

To understand the origin of the high conductivity of SnSe₂, we perform Kelvin Probe Force Microscopy (KPFM) measurements. We prepared a special test structure for KPFM measurement where parallel Au lines are first deposited on a SiO₂/Si substrate, followed by exfoliation of SnSe₂ multi-layer on top, such that the flake partially resides on a metal line, and the rest on SiO₂. An AFM thickness mapping image and the corresponding KPFM mapping of the test structure is shown in Fig. 3a-b. This "metal-bottom/flake-top" structure provides an easy access of the probe tip to the SnSe₂ film and avoids the difficulty of characterization of the SnSe₂ layers hidden underneath the contact metal in a typical "metal-top/flake bottom" structure used in the photoconductor device. The metal lines were grounded during KPFM measurement. The KPFM measurements in Fig. 3c indicate the contact potential difference (CPD) between the tip and the sample, which, in turn, allows us to predict the local work function differences $\Delta W = W_{tip} - W_{sample}$. Here, $W_{tip} = 5.3$ eV. A CPD of 400 mV between tip and SnSe₂ thus suggests that the work function of the SnSe₂ sample is around 4.9 eV. The electron affinity of SnSe₂ has been reported to be around 5.2 eV[15]. Hence, we conclude that the flakes are degenerately n-type doped, explaining high conductivity and low resistance ohmic contacts. Fig. 3c also shows that the work function difference between SnSe₂ and Au is about 150 meV. Note that, the conduction band minimum occurs at the L point (Fig. 3d) in the bulk SnSe₂ Brillouin zone[16], which has a valley degeneracy ($g_v$) of 3. The electron density can be calculated as $n = \int_0^\infty g_v \frac{1}{1+e^{(E-\mu)/k_B T}} \frac{8\pi\sqrt{2}m^{*3/2}\sqrt{E}}{h^3} dE$ where $\mu = 0.3$ eV from KPFM measurement, and $m^* = 0.4 m_0$ [10]. This leads to an estimated electron density in excess of $10^{20}$ cm⁻³. We also observe very weak gate dependence of the output current, in agreement with degenerate doping. The resulting band diagram of a metal-SnSe$_2$-metal (M-S-M) structure is schematically illustrated in Fig. 3e.

## 3.2 Photoresponse of substrate supported and suspended SnSe$_2$ device

The response to a 532-nm wavelength laser spot focused at one of the source edges of the substrate supported IDE device is shown in red in Fig. 2b, indicating strong photoresponse. The suspended device, on the other hand, exhibits a weaker photo response (Fig. 2d). The responsivities ($R = \frac{I_{ph}}{P_{op}} = \frac{I_{tot} - I_{dark}}{P_{op}}$) of two different IDE photoconductors (namely, ID1 and ID2, with similar device dimensions) are plotted in Fig. 4a, as a function of drain voltage ($V_{ds}$). For ID1, we used a blanket exposure of 555 nm green light with power density of 2.4 Wm$^{-2}$. For ID2, we use a white light source, with power density $\approx$ 8.4 Wm$^{-2}$. ID1 exhibits a responsivity of $10^3$ A/W at a $V_{ds}$ of 1 mV, while ID2 achieves $\approx 8.66 \times 10^4$ A/W at 100 mV bias. If $G$ is the gain and $\eta$ is the external quantum efficiency, then we have $R = G \frac{q\eta\lambda}{hc} = G\eta \frac{\lambda \, (nm)}{1243}$. Clearly, the obtained responsivity numbers in Fig. 4 translates to a high gain. The obtained responsivity is also benchmarked against several reports from recent literature on planar photoconductive devices using layered materials. This benchmarking exercise reveals the uniqueness of SnSe$_2$ in terms of achieving large gain at ultra-low operating voltage.

Assuming the shot noise from the dark current of the devices as the primary contributor to the total noise, the specific detectivity ($D^*$) is calculated as $D^* = \frac{R\sqrt{A}}{\sqrt{2qI_{dark}}}$, where $q$ is absolute value of electron charge, and $A$ is the device area. The $D^*$ values, plotted in Fig. 4b, show a monotonic increase with applied bias. At 0.1 V external bias, the IDE device exhibits an impressive specific detectivity of $10^{13}$ Jones.

The measured responsivity of the suspended structure (DS) is $\approx$115 A/W at 100 mV for both types of illumination (532 nm laser and white light), as shown in Fig. 4a. This is lower than the IDE devices by more than two orders of magnitude, but still compares superlatively with, for example, graphene based photoconductors[17][18][19]. The corresponding $D^*$ values of the suspended device also exhibit a similar reduction compared with the substrate supported IDE devices (Fig. 4b).

Owing to the degenerate electron doping in the multi-layer SnSe₂, electrons are the primary current carrying species responsible for photocurrent. Note that the gain $G$ is the ratio of hole trapping time and electron transit time ($G = \frac{\tau}{\tau_{tr}}$). Using $\tau_{tr} = \frac{L^2}{\mu_e V_{ds}}$ we obtain

$$R = \frac{q\eta\lambda}{hc}\left(\frac{\tau}{\tau_{tr}}\right) = \frac{q\lambda\eta\tau\mu_e}{hcL^2}V_{ds} \qquad (1)$$

where $L$ and $\mu_e$ are the length of the device channel and electron mobility, respectively. This explains the strong linearity obtained between $R$ and $V_{ds}$ in Fig. 4 for a given wavelength of excitation. If we take $\eta \approx 1$, using equation (1), the responsivity at 555 nm wavelength translates to $G \approx 2.24 \times 10^3$ at $V_{ds} = 1$ mV for the IDE devices.

### 3.3 Gain Mechanism and role of substrate

To understand the origin of such high gain, we perform scanning photocurrent measurements in an IDE photoconductive device. As schematically illustrated in Fig. 5a, a 532-nm laser is scanned across the device, and the corresponding photocurrent is recorded at $V_{ds} = 1$ mV. The results are summarized in Fig. 5b. We observe a strong non-uniformity in the spatial distribution of the photocurrent with a maximum at the source metal edges, while the minimum occurs inside the channel. Similar observations of non-uniform photocurrent have been made in other layered material based photoconductive devices as well[20]. Strong inherent doping allows efficient band bending at the SnSe₂/metal junction. The strong photocurrent at the source and drain edges can be attributed to a combined effect of band bending induced carrier separation and metal induced hot photo carrier injection. Larger photocurrent when the laser spot is at the source edge compared with the drain edge can be understood by noting that photo-generated electrons are the primary charge carriers in the n-doped channel of the device. When the laser spot is at the drain contact edge, the holes being the minority carriers, recombine with electrons before being collected by the source end.

The high gain suggests that the lifetime of the photo-generated holes is much larger than the electron transit time, that is $\tau \gg \tau_{tr}$. To obtain a quantitative understanding about the trapping time $\tau$, we performed a time dependent laser excitation, as shown in Fig. 5c. The laser spot is focused at the source contact edge, turned on for a while and then turned off. The slow decay of the photocurrent shows an exponential behavior, suggesting slow de-trapping of holes. Once the laser is turned off, the slow decay of the photocurrent (after a sharp fall for a very short period) can be fitted well with single exponential decay expression as $I = I_0 + I_1 e^{-t/\tau_1}$, as shown by the black dashed line in Fig. 5c. Here $I_0$, $I_1$, and $\tau_1$ are fitting parameters. The time constant $\tau_1$ obtained from the fitting is found to be 66 s for this substrate supported photoconductor. This slow process allows sufficient time for successive reinjection of electrons in the system[21], providing high gain. This mechanism is more efficient when the laser spot is at the source metal edge, rather than inside the channel. Due to self-consistent electrostatics, the positive charge of the trapped holes at the source edge forces the bands to be pushed down, in turn, reducing the barrier height existing between the contact and $SnSe_2$[20]. This is schematically shown in Fig. 5d-f. A reduced barrier allows for more efficient injection of electrons from the source into the channel, in turn increasing the photocurrent.

However, this data does not clearly explain whether the holes are trapped in the $SiO_2$ substrate, or confined in the $SnSe_2$ film itself. To segregate the contributions of the substrate and the intrinsic film, we carry out similar transient response in the suspended structure, and the result is summarized in Fig. 6a. Two different decay time scales are clearly observed, and this can be fitted with two different exponential decay rates as $I = I_0 + I_1 e^{-t/\tau_1} + I_2 e^{-t/\tau_2}$, as shown by the black dashed line. The two decay time constants $\tau_1$ and $\tau_2$ are 2.27 s and 53.2 s, respectively. The fast decay with time constant $\tau_1$ immediately after turning off the laser, which is almost absent in the substrate supported device, corresponds to the faster component of the photocurrent, with relatively small gain. The slow decay with time constant $\tau_2$ at the longer time scale corresponds to the slow de-trapping of holes, likely trapped in the defect centers in the $SnSe_2$ film, as schematically depicted in Fig. 6b. The differences in the observations of transient response (Fig. 5c and Fig. 6a) and responsivity (Fig. 4) between the substrate supported and suspended devices indicate that the interfacial traps between $SnSe_2$ and $SiO_2$ substrate play a crucial role in the photoconductive characteristics of substrate supported devices.

The input optical power dependent responsivity of the suspended device is shown in Fig. 6c. Such a decrease of responsivity with increasing optical power can be attributed to the larger photo induced carrier density at higher optical power, resulting in (i) reduced drain field in the $SnSe_2$ layer due to screening, (ii) enhanced recombination of photo-generated electron-hole pairs, and (iii) saturation of trap filling in the SnSe$_2$ film by photo generated holes, suppressing gain.

## 4. Conclusion:

In conclusion, we demonstrate high gain photoconductive devices with impressive responsivity and specific detectivity using inter-digitated electrode structure in multi-layer SnSe$_2$, operating at ultra-low voltages. Such devices will find applications in sensitive photodetection applications, consuming ultra-low electrical power, where speed is not the primary requirement. The gain is found to be suppressed in a suspended device, due to complete decoupling of the channel from the SiO$_2$ substrate, avoiding hole traps. The comparison of gain and transient response between the suspended and substrate supported devices allows us to segregate the substrate effects in photoconductivity. The methodology is applicable to other layered material systems as well to extract the intrinsic photoconductive behavior.

**ACKNOWLEDGMENT**

The authors acknowledge the help of C. Venkatesh in carrying out KPFM measurements. K.M. would like to acknowledge support of a start-up grant from IISc, Bangalore, and the support of grants under Ramanujan Fellowship, Early Career Award, and Nano Mission from Department of Science and Technology, (DST), Government of India.

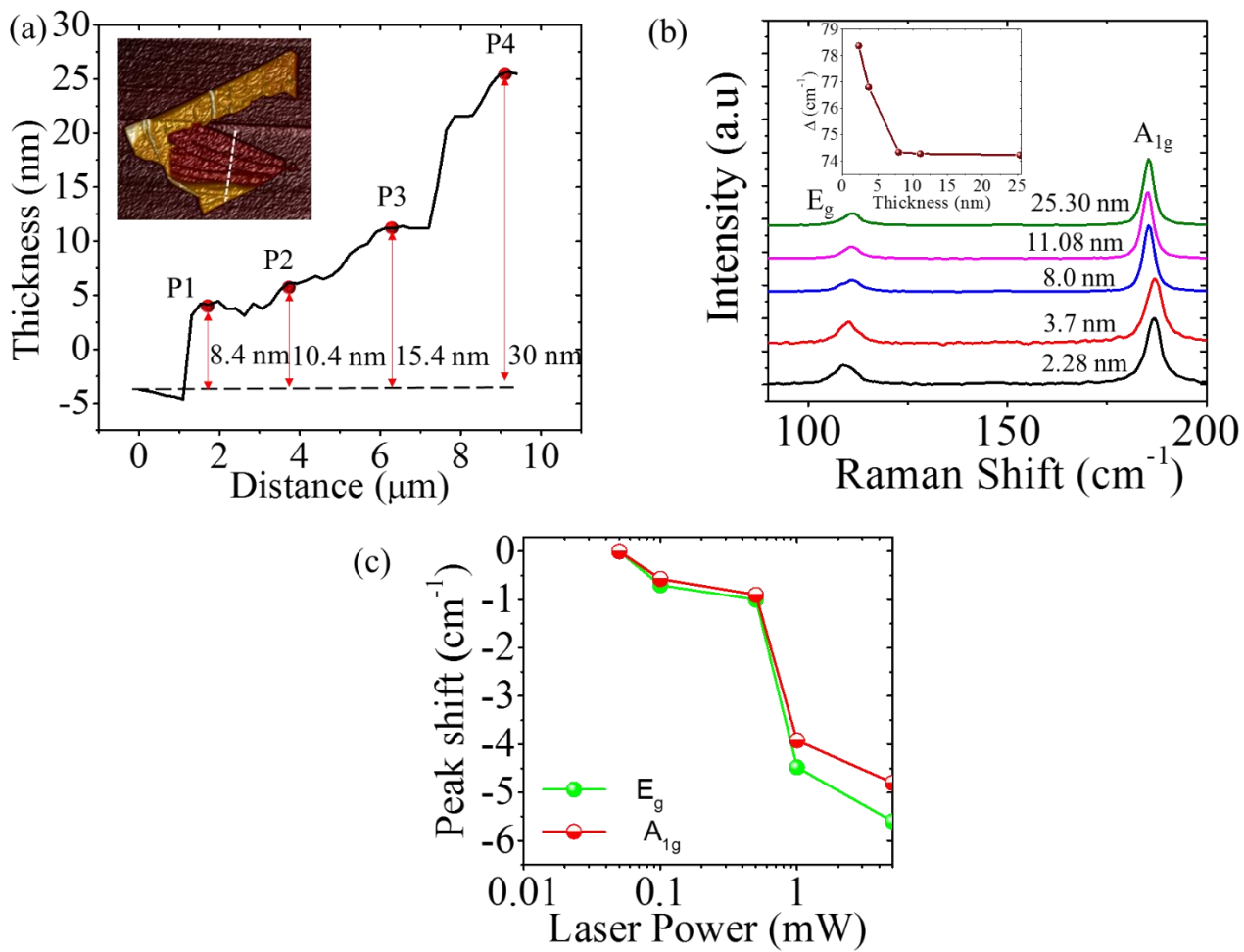

**Figure 1.** (a) Thickness characterization of SnSe$_2$ flakes by AFM. Inset, AFM thickness mapping image of a flake with varying thickness regions. The locus of the scan for the main figure is indicated by white dashed line. (b) Raman shifts for A$_{1g}$ and E$_g$ peaks of SnSe$_2$ flakes with various thickness values. Inset, separation between E$_g$ and A$_{1g}$ as a function of thickness. (c) Raman peak shifts as a function of laser power. Zero corresponds to peak position at lowest laser lower.

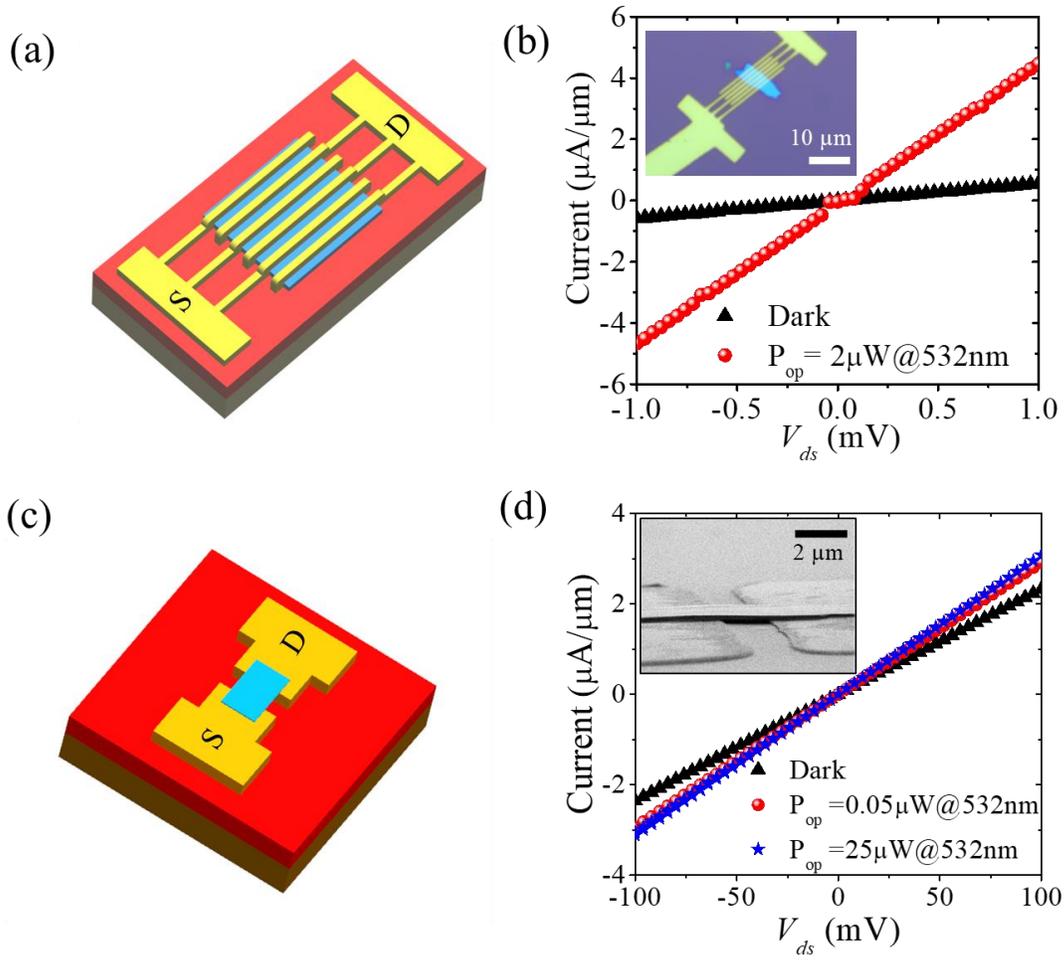

**Figure 2.** (a) Schematic diagram of an inter-digitated electrode (IDE) structure for photoconductive device. (b) Dark current (in black triangle) and current with light (red circles). Inset, Optical image of a fabricated IDE device. (c) Schematic diagram of a suspended device (DS) structure. (b) Dark current (in black triangle) and current with light (red circles and blue stars). Inset, Cross section SEM of a fabricated suspended device.

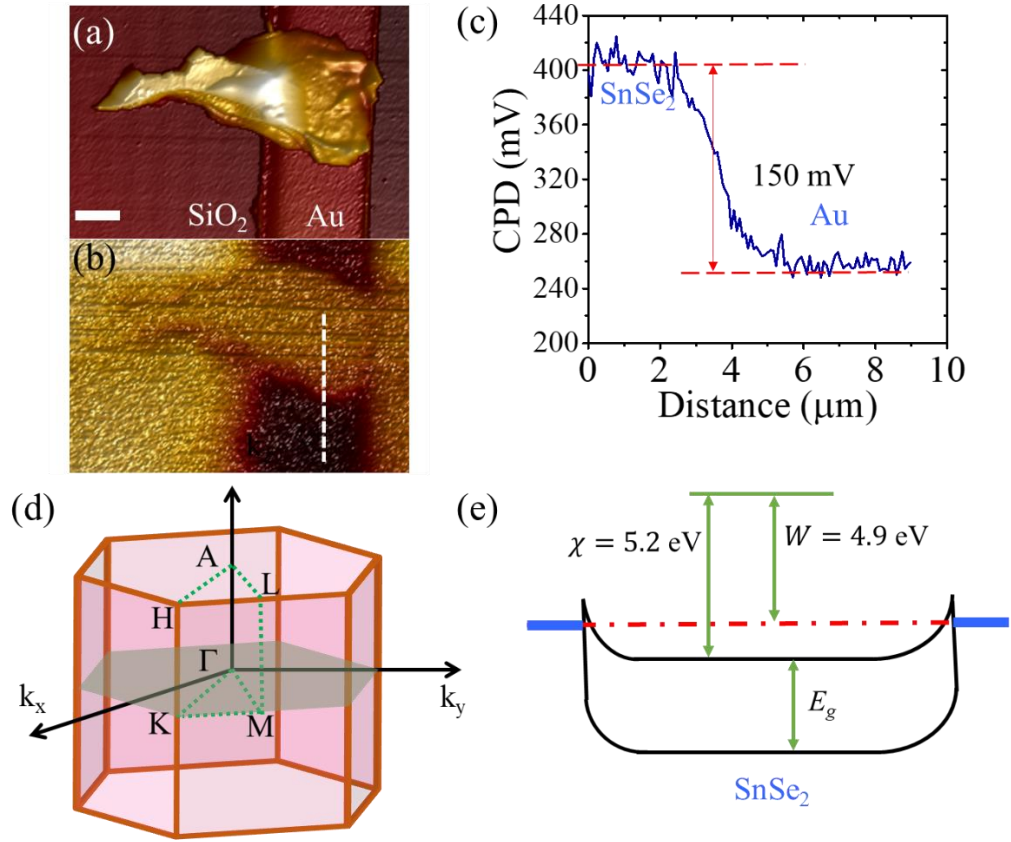

**Figure 3.** (a) AFM thickness scanning image for SnSe$_2$ flake on SiO$_2$ and Au. Scale bar is 2 μm. (b) Contact potential difference (CPD) scanning image in the same region. (c) CPD plotted along the white dashed line in the bottom panel of (a), clearly showing the SnSe$_2$ and Au regions. (d) First Brillouin zone of bulk SnSe$_2$, with the conduction band minimum occurring at L point, with a degeneracy of 3. (e) Schematic band diagram of M-S-M structure for SnSe$_2$ indicating the relative positions of Fermi level and conduction band edge, along with the bandgap.

.

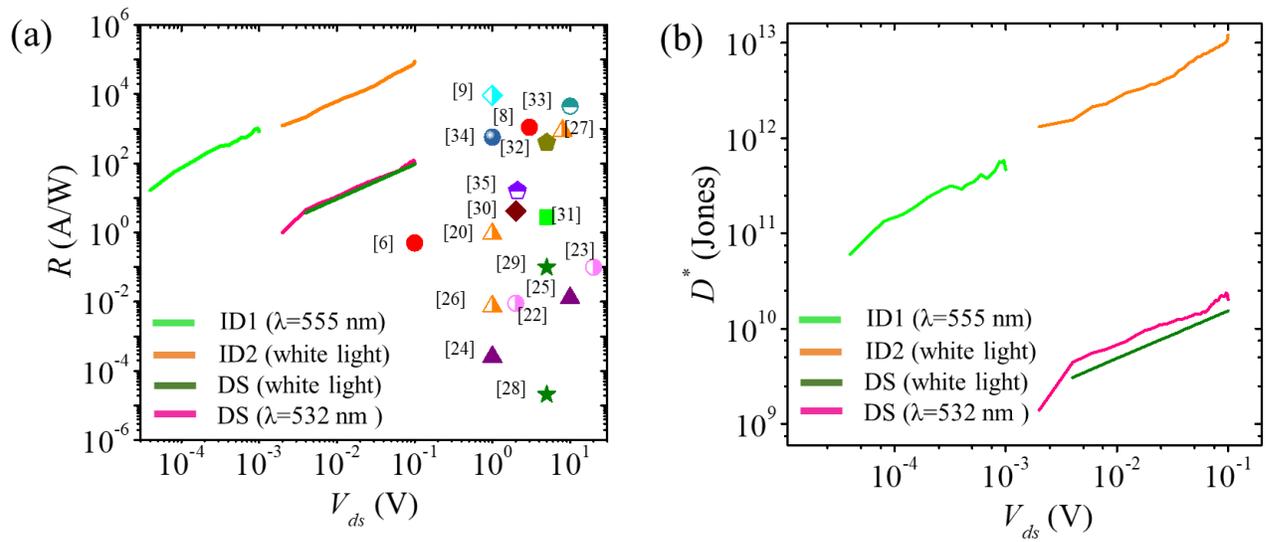

**Figure 4.** (a) Responsivity plotted as a function of applied bias. The data obtained from different devices of this work is shown by solid lines. The data is benchmarked against different planar photoconductive device data from literature (symbols). (b) Specific detectivity extracted from the devices at similar conditions as in (a), plotted as a function of applied bias.

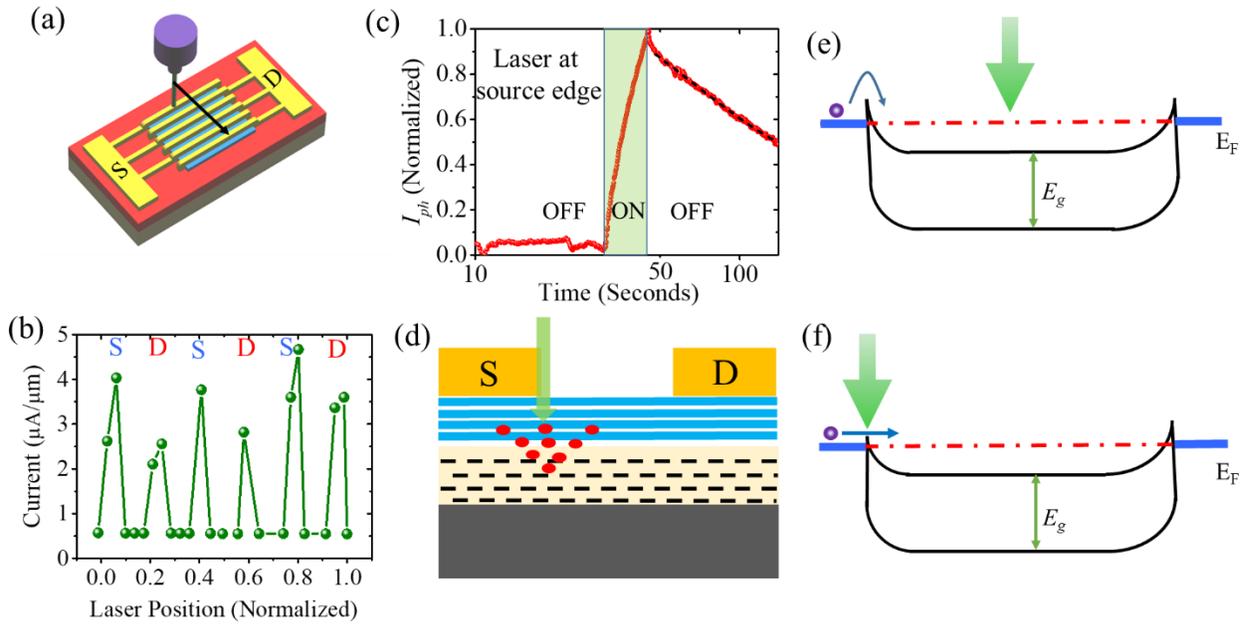

**Figure 5.** (a) Schematic of scanning photocurrent setup. (b) Photocurrent due to a scanning laser in an IDE structure. (c) Time response of the device with the laser spot (wavelength of 532 nm) is focussed at one of the source edges of the IDE. The time axis is in logarithmic scale. The black dashed line shows a fit to the experimental decay with single time constant. (d) Schematic of hole trapping when the laser spot is near the source edge. (e)-(f) The source barrier heights and electron injection from the source are schematically represented depending on the laser position.

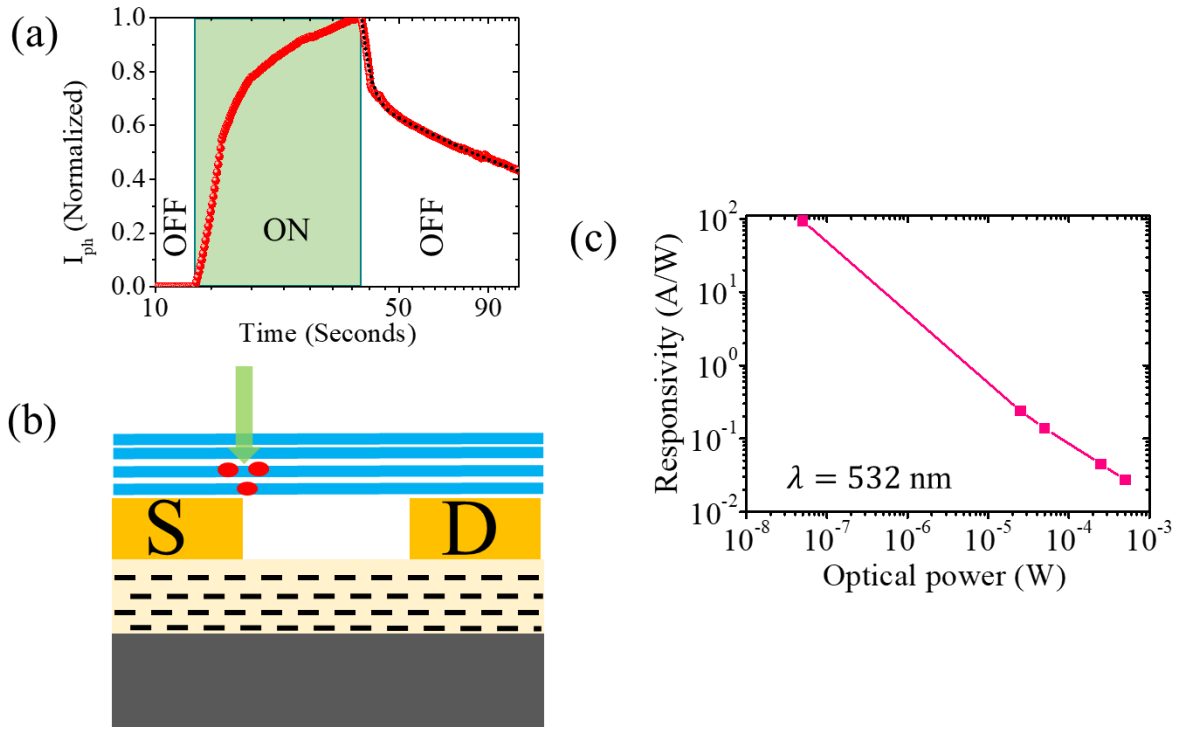

**Figure 6.** (a) Time response of the suspended device with the laser spot (wavelength of 532 nm) being switched on and off. The time axis is in logarithmic scale. The black dashed lines show exponential fitting with two different time constants. (b) Schematic of hole trapping in the suspended device when the laser spot is near the source edge. (c) Responsivity of the suspended SnSe$_2$ device as a function of input power.